\documentclass[structabstract]{aa}  
\usepackage{natbib}
\bibpunct{(}{)}{;}{a}{}{,} % to follow the A&A style
\usepackage{graphicx}
\usepackage{color}
\usepackage{rotating}
\usepackage{txfonts}
\def\mpc{h^{-1} {\rm{Mpc}}}
\def\kms {\rm{km~s^{-1}}}
\begin{document}
\title{Using the large scale quasar clustering to
constrain flat quintessential universes}

\author{Ariel Zandivarez \and H\'ector J. Mart\'\i nez}

\institute{Instituto de Astronom\'ia Te\'orica y Experimental (IATE), CONICET-Observatorio Astron\'omico,
Universidad Nacional de C\'ordoba. Laprida 854, C\'ordoba X5000BGR. Argentina.\\
              \email{arielz@oac.uncor.edu, julian@oac.uncor.edu}
              }

\date{Received XXX, 2008; accepted XXX , 2008 }

% \abstract{}{}{}{}{} 
% 5 {} token are mandatory
 
  \abstract
   {
   }
   {We search for the most suitable set of cosmological parameters that describes
the observable universe. The search includes the possibility of quintessential flat universes, i.e.,
the analysis is restricted to the determination of the dimensionless matter density and the quintessential
parameters, $\Omega_{\rm M}$ and $w_{\rm Q}$, respectively.}
   {Our study is focused on comparing the position of features at large scales in the density fluctuation field 
at different redshifts by analysing the evolution of the quasar two-point correlation function. 
We trace the density field fluctuations at large scales 
using a large and homogeneous sample of quasars ($\sim$ 38000 objects with 
0.3 $\lesssim$ z $\le$ 2.4 and a median $z=1.45$) drawn from the Sloan Digital Sky Survey Data Release Six. 
The analysis relies on the assumption that, in the linear regime, the length scale of a particular feature 
should remain fixed at different times of the universe for the proper cosmological model. 
Our study does not assume any particular comoving length scale at which a feature should be found, but intends 
to perform a comparison for a wide range of scales instead. This is done by quantifying the amount
of overlap among the quasar correlation functions at different times using a cross-correlation technique.
   }
   {The most likely cosmological model is $\Omega_{\rm M}=0.21\pm 0.02$ and 
$w_{\rm Q}=-0.93\pm0.04$, in agreement with previous studies. 
These constraints are the result of a good overall agreement of the correlation function
at different redshifts over scales $\sim 100-300\mpc$.} 
   {Under the assumption of a flat cosmological model, our results indicate that we are living in a 
low density universe with a quintessential parameter greater than the one 
corresponding to a cosmological constant. This work also demonstrates that a large homogeneous 
quasar sample can be used to tighten the constraints upon cosmological parameters.
   }


   \authorrunning{Zandivarez \& Mart\'inez}
   \titlerunning{Constraining flat quintessential universes}

   \maketitle
%

%%%%%%%%%%%%%%%%%%%%%%%%%%%%%%%%%%%%%%%%%%%%%%%%%%%%%%%%%%%%%%%%%%%%%%%
%%%%%%%%%%%%%%%%%%%%%%%%%%%%%%%%%%%%%%%%%%%%%%%%%%%%%%%%%%%%%%%%%%%%%%%%
\section{Introduction} 
Observations of the cosmic microwave background (CMB, e.g. \citealt{spergel07}) 
and type Ia supernovae (e.g. \citealt{riess98,perlm99,astier06})
support the idea that the missing energy in the universe should possess negative 
pressure $p$ and an equation of state $p=w\rho$. One possible candidate for the 
missing energy is the vacuum energy density or cosmological constant $\Lambda$ 
for which $w=-1$ (\citealt{peebles84}).
The resulting cosmological model, $\Lambda$CDM, consists of a mixture of
vacuum energy and cold dark matter. Another possibility is the QCDM cosmology
based on a mixture of quintessence and cold dark matter (\citealt{ratra98}).
The quintessence, which is defined as a fifth element different from baryons,
neutrinos, dark matter and radiation, is a slowly-varying spatially inhomogeneous
component \citep{quinta98} whose $w$ value is less than 0. Many studies restrict 
the range of $w$ to the interval ($-1\le w\le 0$) since this range best fits current
cosmological observations.

The distinction between $\Lambda$CDM and QCDM models is of key importance in
cosmology. Even when both models match the CMB observations well, the QCDM models
seem to be more suitable to describe the universe at high redshifts 
\citep{eisenstein05}. According
to deep redshift surveys, theories should predict a strong large scale clustering
structure and quasar formation at this stage. For this to happen,
the cessation of growth of the structures should occur at earlier
times. Unlike  $\Lambda$CDM, QCDM universes satisfy
this requirement since the larger the values of $w$, the earlier the growth ceases. 
Nevertheless, it is necessary to know how big the differences among
both cosmologies are in order to restrict the possible initial conditions in
the observed universe. To do so, a tuning process of the cosmological 
parameters is needed, combining the best current observations with the most
suitable statistical tools.

Several attempts have been made to tighten the constraints on 
cosmological parameters. Among the tests suggested in the literature we can
mention measurements of the Hubble parameter (e.g., \citealt{wei08,ski08}),
gravitational lensing (e.g., \citealt{zhang07,zhu08,dore07,vaca08}), angular 
sizes of distant objects (e.g., \citealt{daly07,santos08}), gas mass fraction 
in galaxy clusters (e.g., \citealt{chen04,sen08}) and baryon acoustic 
oscillation peaks at large scales (e.g., \citealt{lima07,sapo07,arielsan08}).

In this work, we rely on the fact that if a particular spatial scale can be measured at 
different stages of the evolution of the universe, that scale can then be used 
to determine the values of the cosmological parameters.
At large enough scales ($r>100\mpc$), density perturbations are in the linear
regime and above the present-day turnaround scale. Thus, structures at large
scales, for instance features in the two-point correlation function, 
should be fixed in comoving
length scales at different times in the evolution of the universe.
Since quasars can be observed at very large distances, studying the evolution of 
quasar clustering, i.e., identifying similar features in the density field at different redshifts, 
could restrict the possible universes that match the observations. This approach
has been adopted by \citet{roukema02}, hereafter $R02$, using a relatively small sample of quasars
and finding a wide range of possibilities for the quintessential parameter ($-1.0\le w<-0.5$).

We use a large sample of quasars drawn 
from the Sixth Data Release (DR6; \citealt{dr6}) of the Sloan Digital Sky Survey 
(SDSS; \citealt{sdss}) to infer cosmological parameters. 
Our study relies on the analysis of features at large scales in 
the quasar correlation function. Since there is a large amount of evidence in the 
literature that we are living in a flat universe (e.g. \citealt{arielsan06}), 
we are interested in constraining perfectly flat quintessential universes, i.e., in our 
analysis there are only two parameters to be 
determined: the dimensionless matter density parameter $\Omega_{\rm M}$, 
and $w_{\rm Q}$ ($w$ denoted with the subscript ${\rm Q}$ to specify QCDM cosmologies).
It is important to note that QCDM cosmologies allow a wide spectrum of possibilities 
for the equation of state $w_{\rm Q}$, which could be constant, uniformly evolving 
or oscillatory. In this work, we restrict our analysis to QCDM universes where 
$w_Q$ is constant since there is evidence that this parameter does not evolve in time 
(e.g., \citealt{wang07,gong07}). We follow the line of thought of $R02$
in which they did not assume the scale at which these features should occur, 
but only required consistency of the scale between different times. 

This paper is organised as follows: in Sect.~\ref{sample} we describe the
sample of DR6 quasars; in Sect.~\ref{analysis} we constrain the values of the
cosmological parameters $\Omega_{\rm M}$ and $w_{\rm Q}$, analysing features in the
quasar correlation function at different redshifts; 
we discuss our results in Sect.~\ref{discussion}.
Throughout this paper we assume a Hubble constant $H_0=100~h~\kms~{\rm Mpc}^{-1}$
and all magnitudes are in the AB system. As stated previously, we also assume that
the universe is perfectly flat, i.e, $\Omega_{\rm M}+\Omega_{\rm Q}=1$.
%%%%%%%%%%%%%%%%%%%%%%%%%%%%%%%%%%%%%%%%%%%%%%%%%%%%%%%%%%%%%%%%%%%%%%%%
\begin{figure}
\centering
\includegraphics[width=8cm]{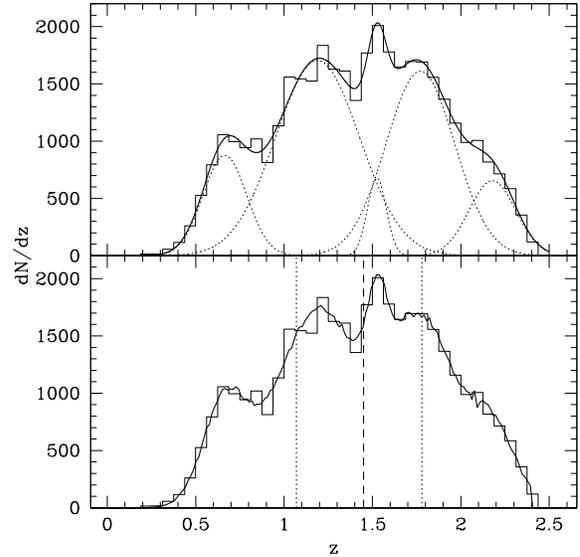}
\caption{Quasar redshift distribution shown as thin line histograms in both panels. 
The upper panel shows in dotted lines the five Gaussian functions whose sum
(thick line) fits the redshift distribution of DR6 quasars. 
The lower panel shows in thick line the redshift distribution
of our random quasar sample obtained by using the fit
shown in the upper panel. 
The vertical dashed line shows the 50th percentile (median value) corresponding to our
two redshift bin analysis, while vertical dotted lines show the 25th (first quartile) and 75th 
(third quartile) percentiles used in the four redshift bins case.}
\label{z}
\end{figure}
%%%%%%%%%%%%%%%%%%%%%%%%%%%%%%%%%%%%%%%%%%%%%%%%%%%%%%%%%%%%%%%%%%%%%%%%
\section{The quasar sample}
\label{sample}
The SDSS DR6 provides the largest homogeneous sample of quasars currently
available. The identification of quasars is achieved with high efficiency and
completeness thanks to the wide-field five band photometric system \citep{gunn98}. 
The high homogeneity of the SDSS spectroscopic sample is accomplished
by selecting the spectroscopic targets
consistently on the basis of their photometric data \citep{richards02,blanton03}.

To ensure further homogeneity in our quasar sample, we select quasars that match the
following criteria \citep{schneider05,ponja05}: 
\begin{itemize}
\item The quasar {\tt primTarget} flag is either "QSO\_CAP" or "QSO\_SKIRT"
or "QSO\_FIRST\_CAP" or "QSO\_FIRST\_SKIRT";
\item $i-$band PSF magnitude corrected for Galactic reddening (using the maps by
\citealt{sch98}) satisfies $15.0\le i\le 19.1$ which are the limits of the 
quasar target selection algorithm;
\item The redshift is $z\le2.4$, since the incompleteness becomes important for higher redshifts;
\item The absolute magnitude is brighter than $M_i=-22.775-5\log(h)$,
computed assuming $\Omega_{\rm M}=0.3$, $w_{\rm Q}=-1.0$ and a 
$K-$correction corresponding to a power-law quasar spectrum with $\alpha=-0.5$;
\item We also exclude the southern Galactic region of SDSS due to differences
in the selection compared to the main survey.
\end{itemize}
Our final quasar sample comprises 38060 quasars with a median redshift 
$z_{\rm med}=1.45$. The redshift distribution of these objects can be seen 
as a solid line histogram in Fig.~\ref{z}.
%%%%%%%%%%%%%%%%%%%%%%%%%%%%%%%%%%%%%%%%%%%%%%%%%%%%%%%%%%%%%%%%%%%%%%%%%%%%
\begin{figure}
\centering
\includegraphics[width=9cm]{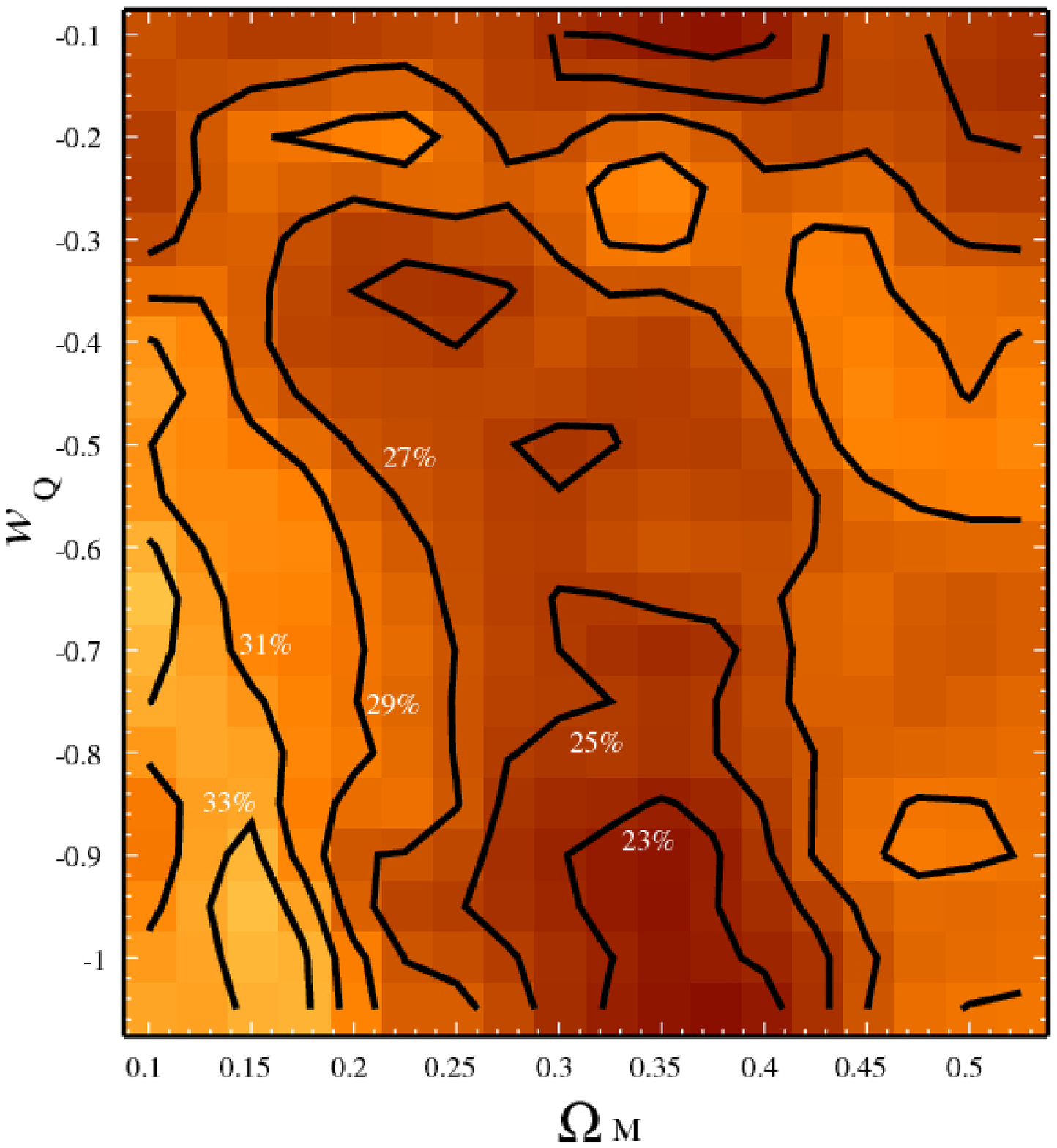}
\includegraphics[width=9cm]{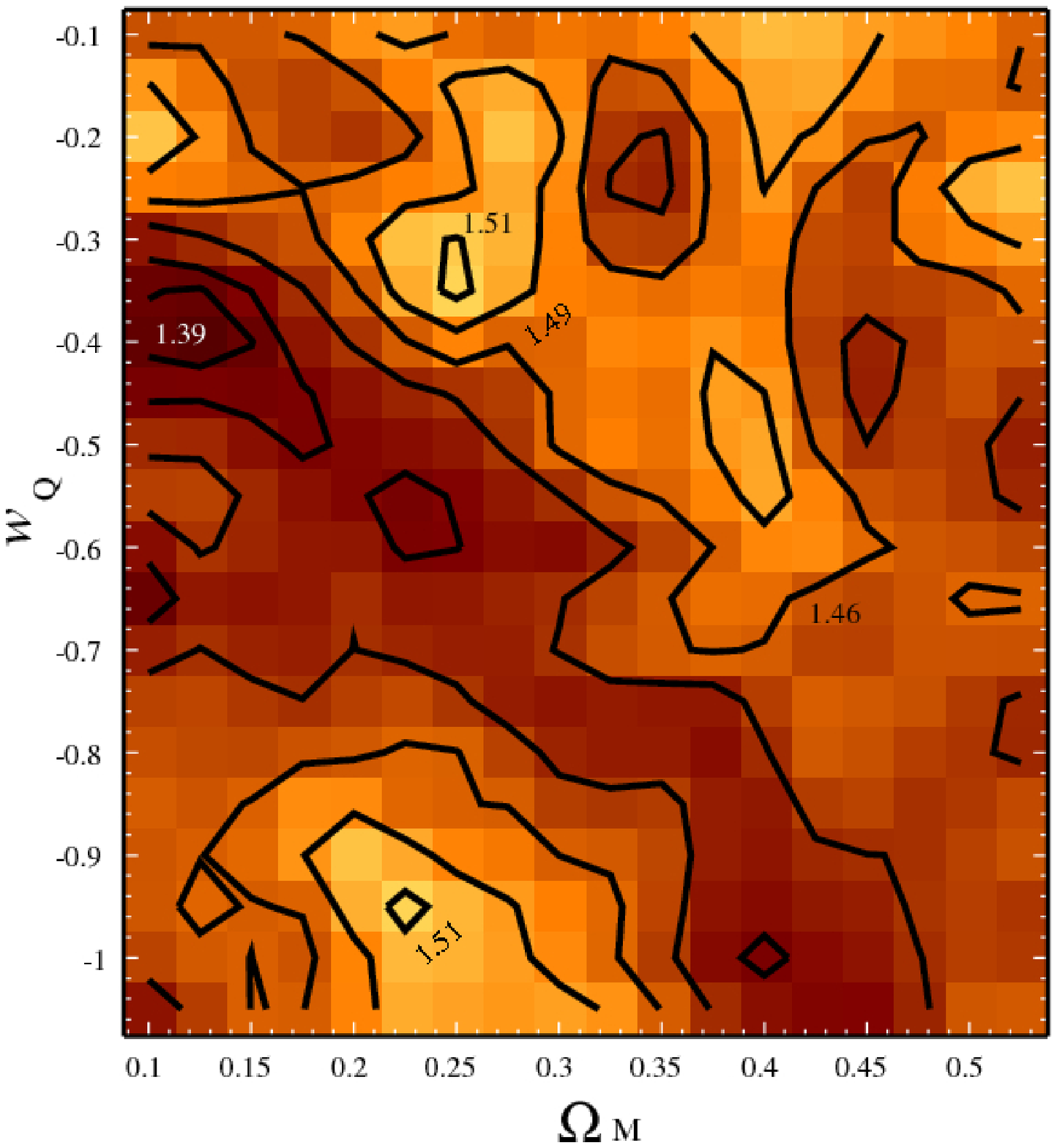}
\caption{$S/N$ analysis for the two redshift bin case.
Upper panel: The percentage of $\xi(r)$ at large scales with $S/N \ge $ 1,
averaged over two redshift bins, i.e., the percentage of scales in $\xi(r)$ we use to
derive cosmological parameters. 
Lower panel: Averaged $S/N$ map for the $\xi(r)$
in two redshift bins including only those scales where $S/N \ge $ 1.
Both panels corresponds to scales $100 \ \mpc \le r \le 300 \ \mpc$ and show
higher values with lighter colours.
}
\label{sn}
\end{figure}
%%%%%%%%%%%%%%%%%%%%%%%%%%%%%%%%%%%%%%%%%%%%%%%%%%%%%%%%%%%%%%%%%%%%%%%%%%%
\section{Determining the parameters $\Omega_{\rm M}$ and $w_{\rm Q}$}
\label{analysis}
The key point of our work is to search for features in the redshift-space two-point correlation
function of quasars that remain fixed in comoving scales as the universe evolves.
We do not attempt to model theoretical quasar correlation functions to compare
with observations since this would require taking into account several not-well-known
factors such as redshift space distortions, quasar bias, survey selection function, 
initial power spectrum, etc.. This might be a hard enterprise since it requires complex
modelling and exhaustive checks with numerical simulations as shown by \citet{arielsan08},
all of them far beyond the scope of this paper. 
Nor do we intend to relate the measured features with scales that resemble
particular events in the history of the universe, such as the sound horizon at recombination
as some authors have already done (e.g. \citealt{guzik07}), since this can be misleading
as pointed out by \citet{arielsan08}.

Our analysis consists of comparing the positions of features (i.e. peaks and valleys) 
in the linear regime of the quasar two-point correlation function, $\xi(r)$, at different redshifts.
We are interested in a comparison among $\xi(r)$s at different times over a wide
range of scales, and not just finding an agreement on a particular 
comoving scale as has been done in previous works (e.g. $R02$).
Regarding the number of redshift bins in which the quasar sample will be split,
our approach is twofold: first, we divide the sample into two equal number redshift bins;
second, we study the possible effects of evolution in our results by dividing them into four redshift bins.
The first choice aims to construct samples of quasars as large as possible to
allow for a fair detection of features in the $\xi(r)$ with 
signal-to-noise ($S/N$) ratios as high as possible for a wide range of comoving scales. 
Nevertheless, it should be noticed that the size of the redshift bins could allow for evolution of
the quasar population within a single bin and thus bias the results. Therefore, 
we intend to understand possible evolution biases on the results that we found in the two redshift bin case 
by means of a four redshift bin analysis, even though this case will be affected by larger errors 
bars in the $\xi(r)$.

\subsection{Computing the quasar correlation function}
In order to keep the noise as low as possible, we calculate the comoving separations directly
in three-dimensional, flat, comoving space, assuming a perfectly homogeneous metric.
That is, we do not attempt the commonly used procedure that consists of deriving the 
spatial two-point correlation function from the projected correlation function since that 
would require us to split pairs into a two-dimensional array, which would increase shot noise.

The cosmology is implicitly involved in the spatial two-point correlation function through
the computation of the comoving distances. The comoving distance-redshift relation 
in a quintessence flat universe is given by:
\begin{equation}
r(z)=\frac{c}{H_0}\int^{z}_0\frac{dz}
{\sqrt{\Omega_{\rm M}(1+z)^3+\Omega_{\rm Q}(1+z)^{3(1+w_{\rm Q})}}},
\label{rz}
\end{equation}
where $\Omega_{\rm Q}=1-\Omega_{\rm M}$ and $w_{\rm Q}$ is constant. 
We explore different cosmologies, varying the parameters $0.1< \Omega_{\rm M} \leq 0.55$
and $-1.1\leq w_{\rm Q} \leq -0.1$ in steps of $\Delta \Omega_{\rm M}=0.025$ and 
$\Delta w_{\rm Q}=0.05$. We are not interested in higher $\Omega_{\rm M}$ values since there is compelling
evidence that we are living in a low matter density universe (e.g. \citealt{arielsan06,spergel07}).

%%%%%%%%%%%%%%%%%%%%%%%%%%%%%%%%%%%%%%%%%%%%%%%%%%%%%%%%%%%%%%%%%%%%%%%%%%%%
\begin{figure*}
\centering
{\includegraphics[width=8cm]{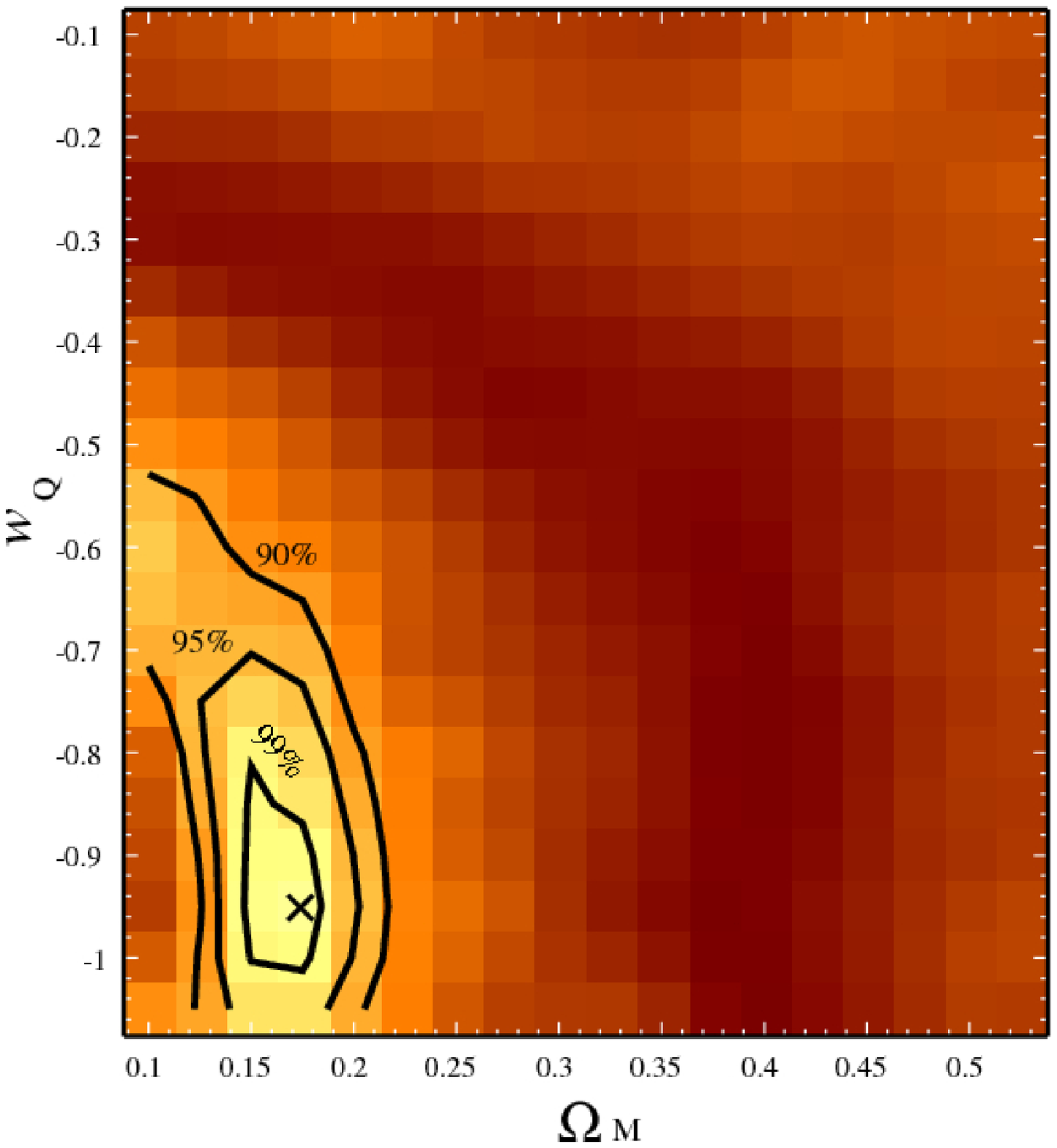}\includegraphics[width=8cm]{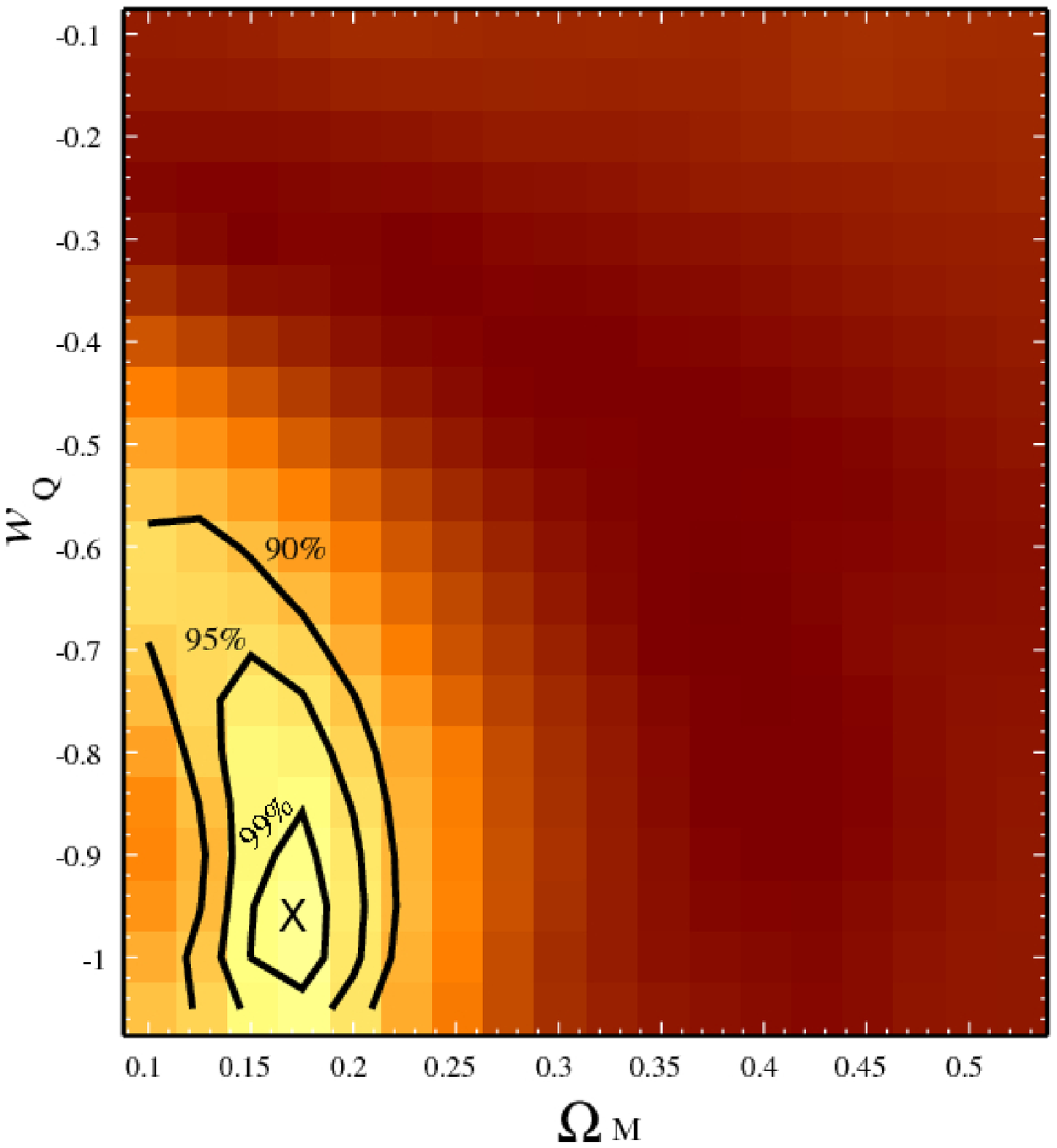}}
{\includegraphics[width=8cm]{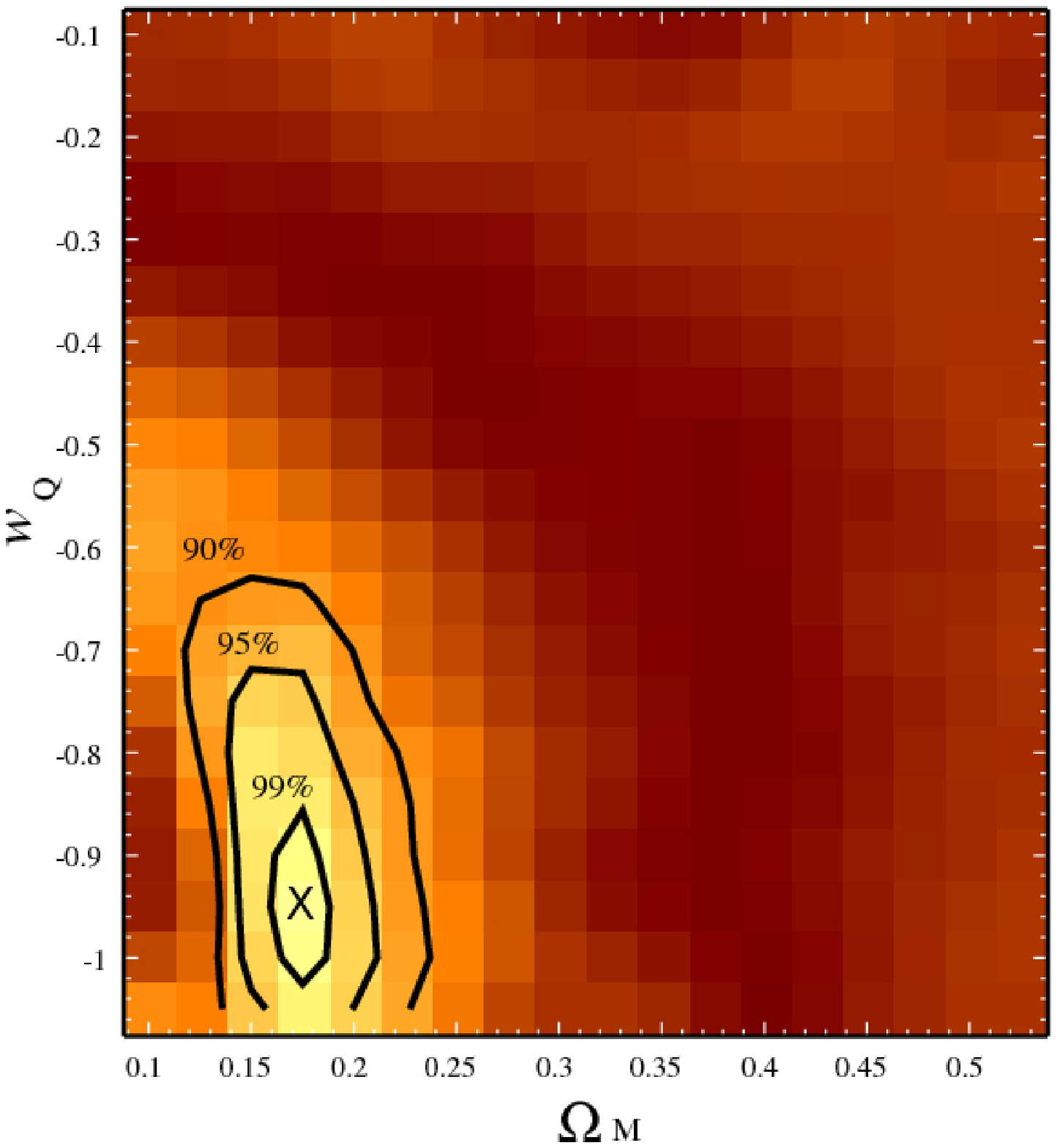}\includegraphics[width=8cm]{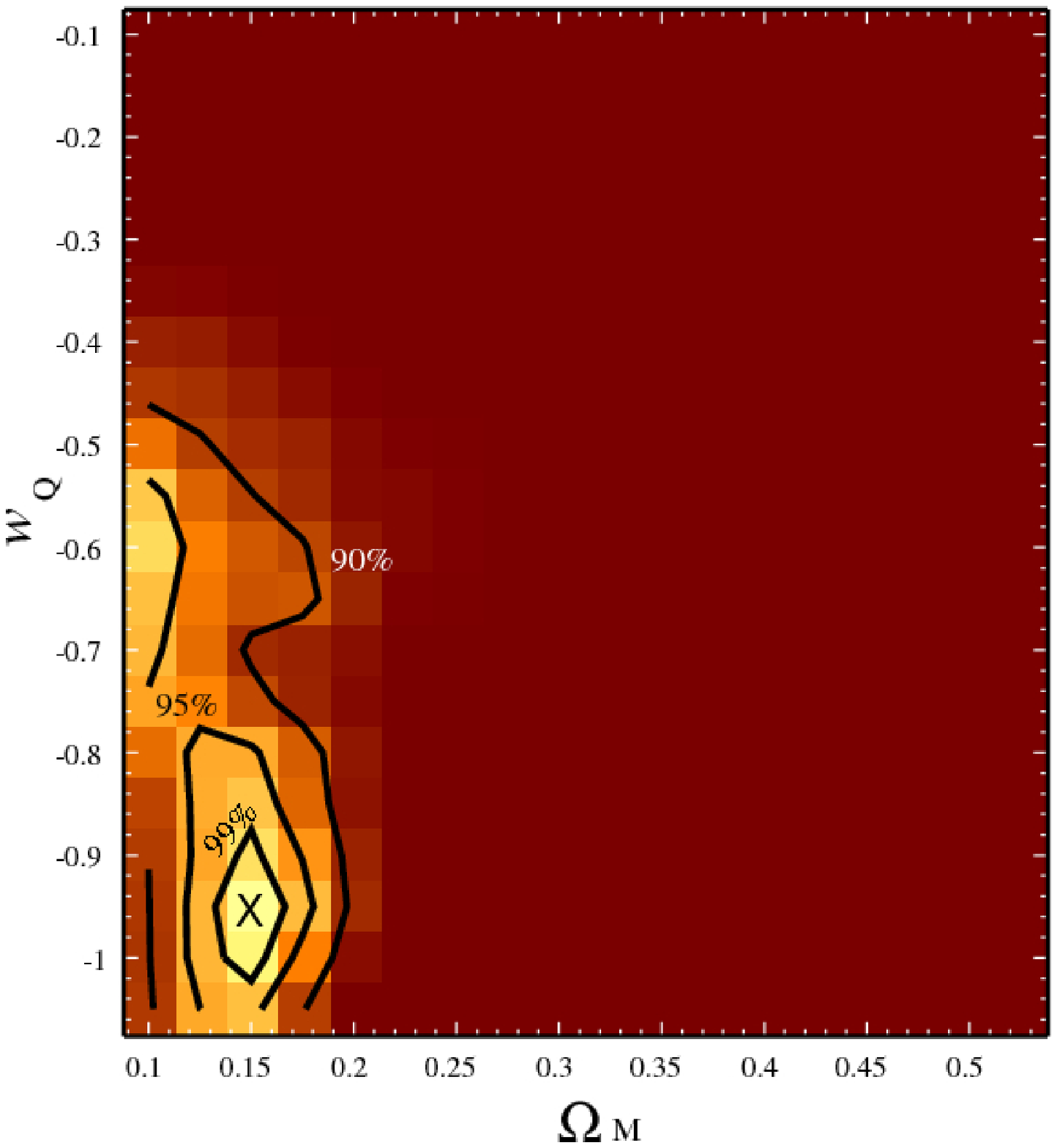}}
\caption{($\Omega_{\rm M}-w_{\rm Q}$) maps for the "{\it cross-correlation}" ($CC^{(2z)}$) of 
the $\xi(r)$ in two redshift bins. The upper left panel shows $CC^{(2z)}_{[1,2]}$ taking into account peaks and
valleys in the $\xi(r)$, while the lower panels show the $CC^{(2z)}_{[1,2]}$ only for peaks (left) 
and valleys (right).
The cross-correlation in these panels were computed only for $S/N \ge $ 1, except for the upper right
panel, where $CC^{(2z)}_{[1,2]}$ is shown taking into account peaks and valleys without any $S/N$ restriction. 
Higher values of the cross-correlation correspond to lighter colours. Labels in contour levels show 
the percentage of the map outside the contours.
}
\label{cxi}
\end{figure*}
%%%%%%%%%%%%%%%%%%%%%%%%%%%%%%%%%%%%%%%%%%%%%%%%%%%%%%%%%%%%%%%%%%%%%%%%%%%

For each pair of parameters, $\Omega_{\rm M}$ and $w_{\rm Q}$, we assign comoving positions
of quasars by using their redshifts in Eq.~\ref{rz} and compute the
correlation functions according to \citet{ls93}:
\begin{equation}
\xi(r)=\frac{DD(r)-2DR(r)+RR(r)}{RR(r)},
\end{equation} 
where $DD(r)$, $DR(r)$ and $RR(r)$ are the normalised number of quasar-quasar,
quasar-random and random-random pairs with comoving separations within $r\pm\Delta r/2$.
For the purposes of this work we use linear bins in comoving distances
with $\Delta r=5\mpc$. 

The random sample was constructed to have the same angular coverage of the survey and 
a smooth redshift distribution that matches that of the quasars. 
The angular mask was constructed using routines included in the 
HEALPix\footnote[1]{Hierarchical Equal Area iso-Latitude Pixelization, \citet{healpix}} 
package to pixelize the sky coverage of the SDSS DR6. As regards the redshift distribution 
of the random points, we have found that a remarkably good fit to the redshift distribution of
quasars is the sum of five Gaussian functions as shown in the upper panel of Fig.~\ref{z}. 
By fitting this function we have a smooth redshift distribution
that closely follows that of the quasars. We then draw the
redshifts of the individual random points from this distribution and obtain the
distribution shown in the lower panel of Fig.~\ref{z}.
The random sample is 20 times denser than the sample of quasars in order to 
reduce shot noise.

For the comoving scales that we are dealing with ($r \ > \ 100 \ \mpc$), 
we compute $\xi(r)$ error bars using the bootstrap technique. 
This method becomes an unbiased estimate of the ensemble error when the number of 
pair counts is large compared to the numbers of objects in the sample (\citealt{boot1,boot2}).
Hence, for each correlation function, we also have 20 bootstrap estimates. 

In order to avoid spurious features, we smooth the correlation 
function and each bootstrap estimation using a Gaussian filter
with $\sigma_{\rm smooth}=5\mpc$. It is worth mentioning that smoothing the 
pair counts instead of the correlation function produces comparable results.
We use these smoothed correlation functions to extract cosmological information 
and the smoothed bootstrap estimates to compute error bars.
%%%%%%%%%%%%%%%%%%%%%%%%%%%%%%%%%%%%%%%%%%%%%%%%%%%%%%%%%%%%%%%%%%%%%%%%%%%%%%%%%%%%%%%%
\begin{figure}
\centering
\includegraphics[width=8.5cm]{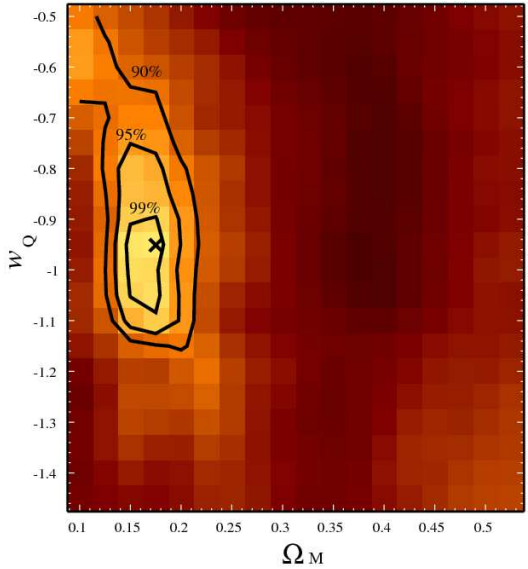}
\caption{
($\Omega_{\rm M}-w_{\rm Q}$) map shifted to lower values of $w_{\rm Q}$.
The  map shows $CC^{(2z)}$ of the $\xi(r)$ for two redshift bins.
The minimum comoving distance of the map is $r_{min}=$110 $\mpc$. Each model is computed only for $S/N\ge $ 1.  
Labels in contour levels show the percentage of the map outside the contours.
}
\label{corr}
\end{figure}
%%%%%%%%%%%%%%%%%%%%%%%%%%%%%%%%%%%%%%%%%%%%%%%%%%%%%%%%%%%%%%%%%%%%%%%%%%%%%%%%%%%%%%%%

%%%%%%%%%%%%%%%%%%%%%%%%%%%%%%%%%%%%%%%%%%%%%%%%%%%%%%%%%%%%%%%%%%%%%%%%%%%%
\begin{figure}
\centering
\includegraphics[width=9cm]{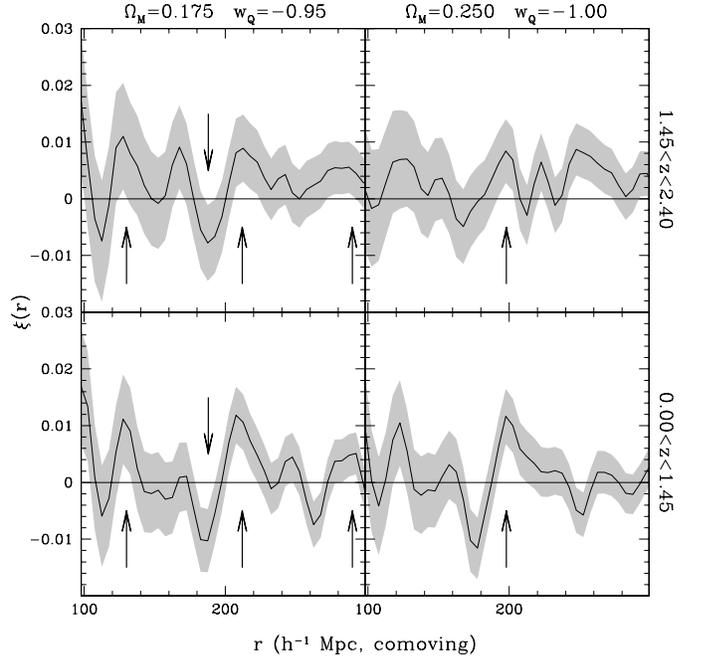}
\caption{The two point correlation functions for the preferred (left panels)
and standard (right panels) cosmological models. Upper and lower panels display the $\xi(r)$s
for the highest and lowest redshift bins respectively. Error bars were computed using the bootstrap
re-sampling technique and are shown as a grey region. Up and down arrows point out features with
$S/N\ge 1$ that remain fixed in comoving scales for both redshift bins.
}
\label{xi2}
\end{figure}
%%%%%%%%%%%%%%%%%%%%%%%%%%%%%%%%%%%%%%%%%%%%%%%%%%%%%%%%%%%%%%%%%%%%%%%%%%

\subsection{Estimating $\Omega_{\rm M}$ and $w_{\rm Q}$}

As it has been stated in previous works (e.g. \citealt{roukema01},~$R02$), 
features at large comoving scales should remain unchanged over time for the proper
cosmology. Our purpose is to extract as much information as possible from the 
features measured in the $\xi(r)$s at different redshifts. 
To achieve this, we do a full scale comparison among $\xi(r)$s. 
This approach should introduce tighter constraints on the cosmological models.

\subsubsection{The statistical tool}

To quantify the goodness of each cosmological model pair ($\Omega_{\rm M}$,~\ $w_{\rm Q}$),
we use a function that allows us to check for similarities among $\xi(r)$s at 
different redshifts. This function is the {\it cross correlation} and it is an integral
that expresses the amount of overlap between two functions. Formally, the cross correlation of 
two real functions $f$ and $g$ is defined as
\begin{equation}
 (f \star g) (\tau) = \int_{-\infty}^{\infty} f(t) \ g(\tau+t) \ dt
\end{equation}
where $\tau$ is a lag applied to $g$. Particularly, we are interested in the case of 
a lag $\tau=0$. Hence, the discrete formula for the cross correlation in our study is
\begin{equation}
 CC_{[i,j]} = (\xi_i \star \xi_{j}) = \sum_{r=r_{min}}^{r_{max}} \xi_i(r) \ \xi_{j}(r) 
\label{cc}
\end{equation}
where $\xi$ is the two-point correlation function and the subscripts $i$ and $j$ denote 
particular redshift bins.
It should be taken into account that we have restricted the sum over a finite range of 
comoving distances ($r_{min}$, $r_{max}$) at which we perform our analysis of the $\xi(r)$s.
Particularly, we choose as a maximum distance $r_{max}=300 \ \mpc$ since larger scales
have very low $S/N$, while $r_{min}$ is used as a variable taking values $\ge 100 \ \mpc$.

Since we are interested in considering only the features 
that show a true signal, from now on (unless specified otherwise) we only use those points
in $\xi(r)$ whose $S/N$ ratio is greater than unity. Points that do not 
exceed this threshold are set to zero in the computation of the quantity $CC$. 

\subsubsection{The two redshift bin case}

Our first approach is to split the quasar redshift distribution into two equal number
subsamples. These subsamples are called the low redshift subsample, $0< z \leq1.45$, and the high
redshift one, $1.45< z \leq2.4$ with median redshifts of 1.07 and 1.78 respectively.
The purpose of this choice is to produce large samples of quasars that allow the measurement of
many statistically significant features.

%%%%%%%%%%%%%%%%%%%%%%%%%%%%%%%%%%%%%%%%%%%%%%%%%%%%%%%%%%%%%%%%%%%%%%%%%%
\begin{figure*}
\centering
{\includegraphics[width=8cm]{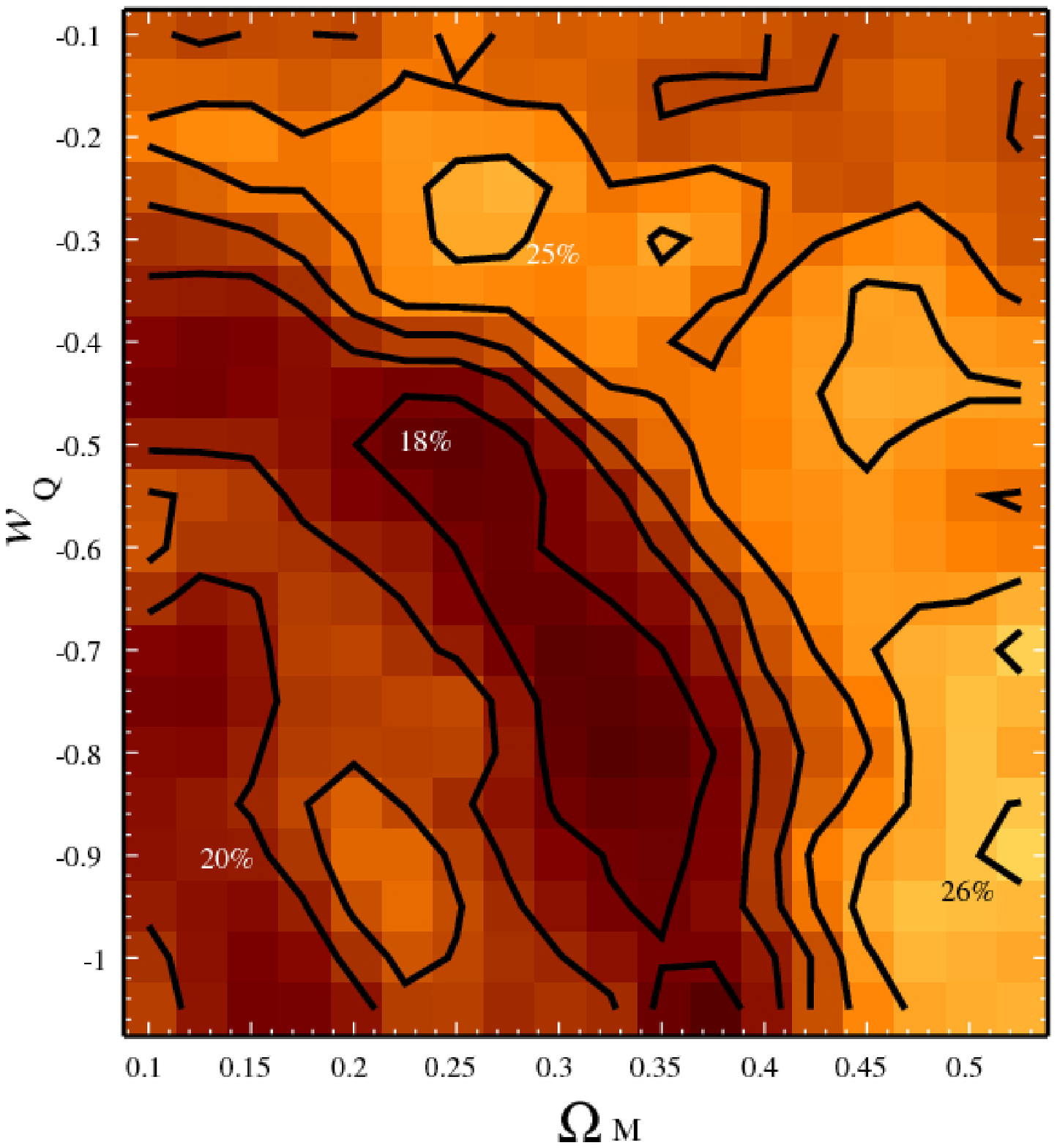}\includegraphics[width=8cm]{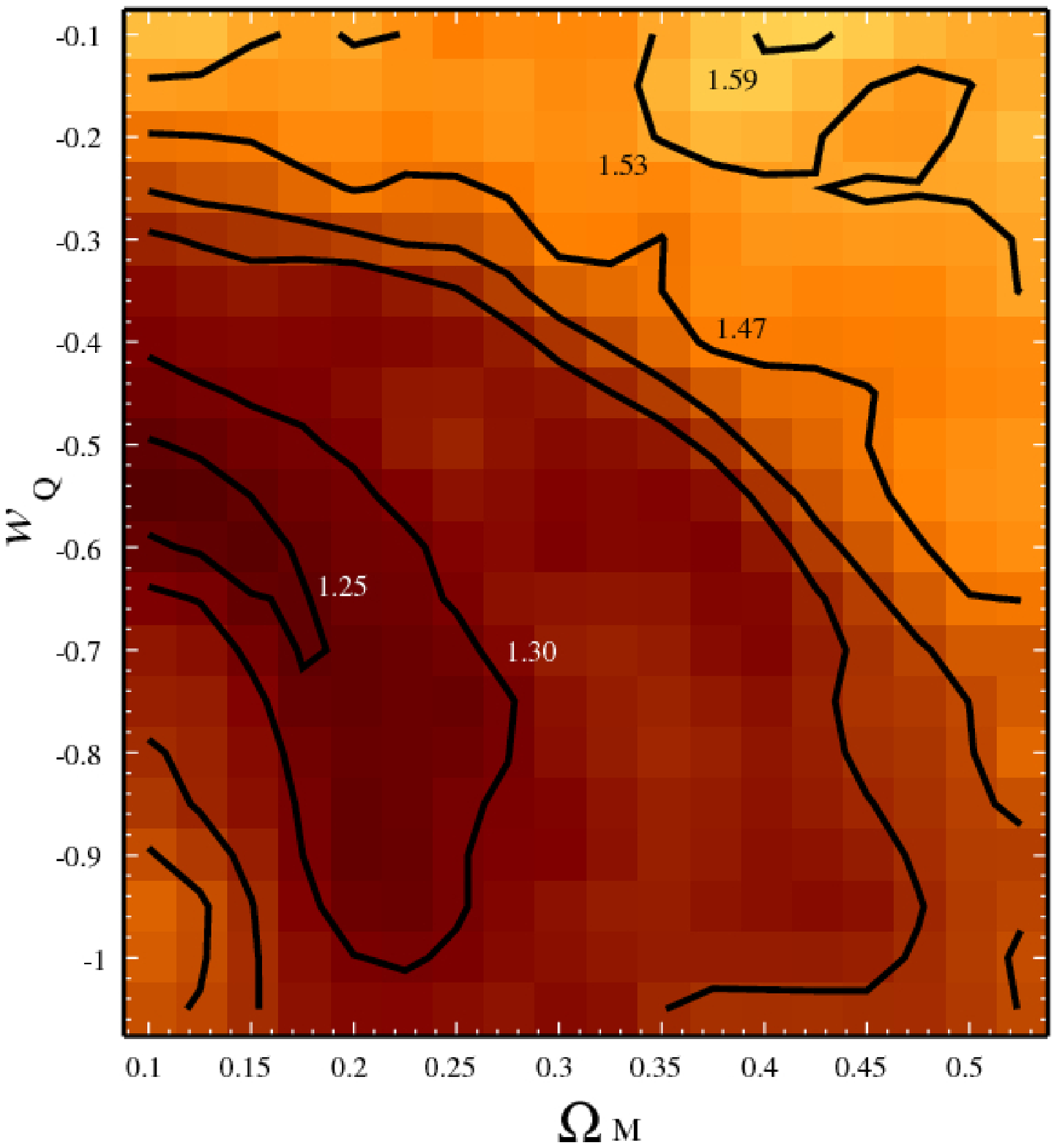}}
{\includegraphics[width=8cm]{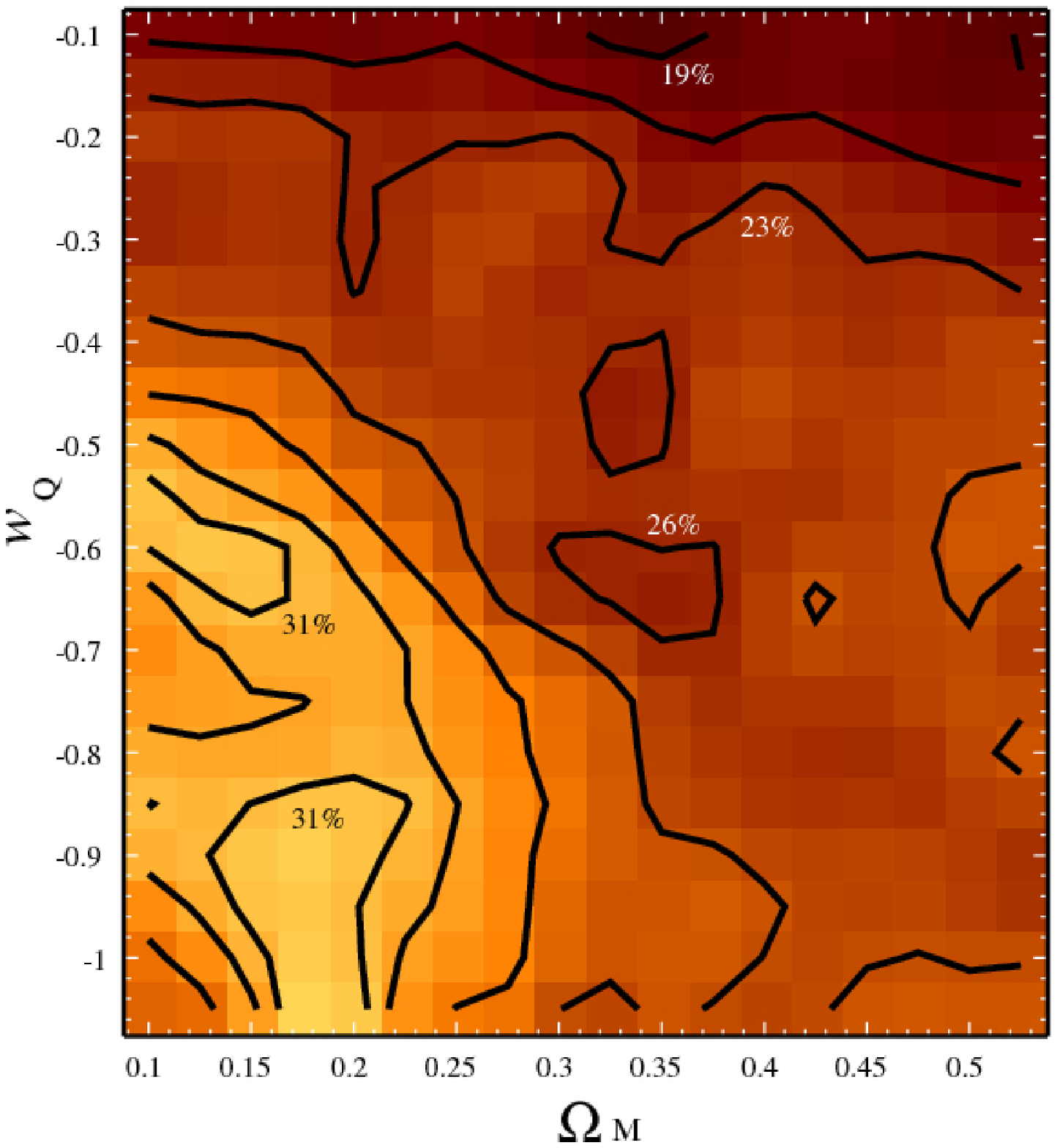}\includegraphics[width=8cm]{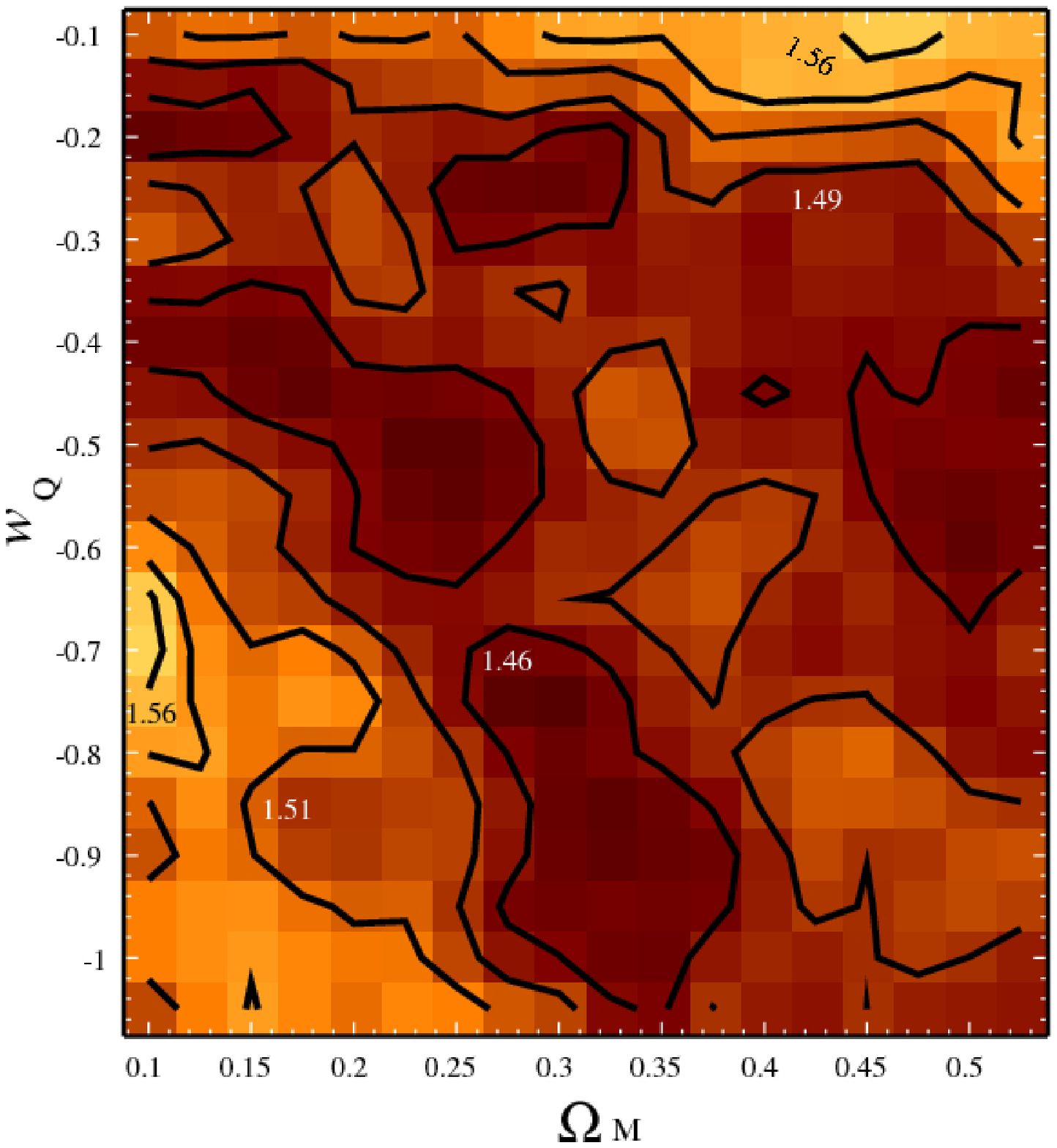}}
\caption{$S/N$ analysis for the four redshift bin case.
Left panels: The percentage of the $\xi(r)$ at large scales with $S/N \ge $ 1, averaged over
the redshift bins 1-3 (upper panel) and 2-4 (lower panel). Right panels: Averaged $S/N$
map for the $\xi(r)$ for redshift bins 1 and 3 (upper panel) and 2 and 4 (lower panel) including
only those scales where $S/N \ge $ 1. All panels correspond to scales $100 \ \mpc \le r \le 300 \ \mpc$
and show higher values with lighter colours.
}
\label{sn3}
\end{figure*}
%%%%%%%%%%%%%%%%%%%%%%%%%%%%%%%%%%%%%%%%%%%%%%%%%%%%%%%%%%%%%%%%%%%%%%%%%%%

Which cosmologies are the most likely to produce significant features in
$\xi(r)$ ? In order to answer this question we construct two significance maps which are 
shown in Fig.~\ref{sn}. The first map (upper panel) shows the percentage of scales in the $\xi(r)$
that shows $S/N \ge 1$ (significant features) averaged over the two redshift bins, while the second one 
(lower panel) shows the averaged $S/N$ for the $\xi(r)$ for two redshift bins, also using significant
features only. The $\xi(r)$s are restricted to the range $100\mpc \le r \le 300\mpc$.
In both panels of Fig.~\ref{sn} lighter colours indicate higher values.   
From the first map we observe that our restriction on $S/N$ implies that the useful range of scales
on the $\xi(r)$, on average, runs from 23\% to 33\%. On the other hand, the second map shows that
the averaged $S/N$ lies in the range 1.39 to 1.51.
From the contour levels and colours in this Fig. we can observe that there are many regions in both
maps that are prone to a significant range of scales that are useful and a relatively high
average $S/N$.
Beyond that, an important issue that should be noticed is that all cosmological 
pairs on the map have useful scales in the $\xi(r)$ and therefore
all models are plausible in showing coincident features at different times.

To quantify the level of agreement among $\xi(r)$s, we use the cross correlation 
denoted by $CC_{[1,2]}^{(2z)}$, where we have added an upper-script, $(2z)$, denoting the
current case. This statistic has the capability of enhancing features that 
occur at similar scales while erasing those that do not appear at the same distance.
Moreover, since there is a sum over a wide range of comoving scales, several 
coincident signals for a particular cosmological pair produce higher values of $CC$ 
implying a better match between the $\xi(r)$s at different redshifts.

The $CC_{[1,2]}^{(2z)}$ maps with $r_{min}=110 \ \mpc$ are shown in Fig.~\ref{cxi}.
All panels show contour levels around the $CC_{[1,2]}^{(2z)}$ maxima that indicate 
the percentage of the map outside their boundaries. 
The upper left panel shows the $CC_{[1,2]}^{(2z)}$ computed taking into account all
features (peaks and valleys) in the $\xi(r)$s. From this panel, we can see 
that the most likely cosmological models are in a small region in the lower left corner of the
map. The black cross shows the maximum $CC_{[1,2]}^{(2z)}$ value that corresponds to 
$\Omega_{\rm M}=0.175$ and $w_{\rm Q}=-0.95$. 

Since this result is achieved by using all features in the $\xi(r)$s,
an interesting question is whether a particular feature (peaks or valleys) contributes to a particular
region in the map.
This issue can be answered by computing the cross correlation statistics using eq.~\ref{cc},
but taking into account positive (peaks) or negative (valleys) significant ($S/N \ge \ 1$) 
values of the $\xi(r)$ only. 
The resulting maps can be seen on the lower left (peaks only) and lower right (valleys only)
panels in Fig.~\ref{cxi}. 

From the left panel, we can see that peaks restrict the 
area displayed by the contour levels compared with the ones obtained with all features. 
The contours are reduced, ruling out some cosmological pairs with $w_{\rm Q}\ge -0.6$
and are very slightly shifted to the right (higher values of $\Omega_{\rm M}$). Even so,
the preferred cosmological pair remains unchanged.
On the other hand, valleys alone give lower $\Omega_{\rm M}$ values 
($\sim 0.150$) and produce an elongation of contour levels towards higher values of $w_{\rm Q}$. 
Nevertheless, their contribution to the combined feature analysis is not strong enough 
to prevail over the signal preferred by peaks.

Finally, it is interesting to know what would be the effect of relaxing the restriction 
imposed on the $S/N$ of the features. In the upper right panel of Fig.~\ref{cxi} we
construct the $CC_{[1,2]}^{(2z)}$ map without imposing any value on the $S/N$ of the $\xi(r)$s.
From the colour intensity of the map it is clear that the main 
effect (compared to the upper left panel) is the widening of the area of preferred cosmologies. 
Even when this map enhances the importance of some cosmological models that are not statistically
significant, the resulting preferred cosmology turns out to be very similar to the one
obtained from significant features only.

Since our result on $w_{\rm Q}$ is close to the limit of the $w_{\rm Q}$ range used so far,
it would be interesting to test whether the exclusion of $w_{\rm Q} < -1.1$ is biasing our results.
Therefore, we repeated the cross correlation analysis over a shifted range, 
$-1.45\le w_{\rm Q}\le-0.50$. As can be seen from Fig.~\ref{corr} our results are stable, hence
no bias arises from our original choice for the $w_{\rm Q}$ range.

To illustrate the ability of our method to compare $\xi(r)$s at different times, in
Fig.~\ref{xi2} we show a comparison between the best cosmological pair obtained using
the $CC_{[1,2]}^{(2z)}$ map ($\Omega_{\rm M}=0.175$ and $w_{\rm Q}=-0.95$, left panels) and the
standard model ($\Lambda CDM$, $\Omega_{\rm M}=0.250$ and
$w_{\rm Q}=-1.00$, right panels). The grey region in each panel shows the $1\sigma$
bootstrap errors bars.
The best model is clearly better than the standard one given 
the number of coincident significant features at different scales. The arrows in this
Fig. show features observed at the same comoving scale that satisfy
$S/N \ge \ 1$. While the preferred cosmological model obtained from
our maps has four coincident features, we can only observe one matching feature for the
standard cosmological model.

We have also checked whether the resulting best model depends on the choice of $r_{min}$
by computing $CC_{[1,2]}^{(2z)}$ maps varying $r_{min}$ from 120 to 190 $\mpc$. 
In all cases we found that the preferred pair of cosmological parameters is the same obtained 
for $r_{min}=110 \ \mpc$, indicating the stability of our results.

%%%%%%%%%%%%%%%%%%%%%%%%%%%%%%%%%%%%%%%%%%%%%%%%%%%%%%%%%%%%%%%%%%%%%%%%%%%
%%%%%%%%%%%%%%%%%%%%%%%%%%%%%%%%%%%%%%%%%%%%%%%%%%%%%%%%%%%%%%%%%%%%%%%%%%%
%%%%%%%%%%%%%%%%%%%%%%%%%%%%%%%%%%%%%%%%%%%%%%%%%%%%%%%%%%%%%%%%%%%%%%%%%%%

\subsubsection{The four redshift bin case}

Given the size of the redshift bins used in the previous case, the evolution of the quasar 
population within each bin might be biasing the results. In this subsection, we
use four redshift bins to avoid too much evolution within a single bin.
This analysis tests the stability of the results obtained with two redshift bins
even when narrower redshift bins imply larger errors in the $\xi(r)$.     
We divide the sample into four equal number redshift bins:
\begin{enumerate}
\item $0.00< z \leq1.07$ 
\item $1.07< z \leq1.45$
\item $1.45< z \leq1.78$ 
\item $1.78< z \leq2.40$.
\end{enumerate}
With these subsamples we choose to calculate two estimates of the cosmological 
parameters using the cross correlations between redshift bins 1-3 and 2-4.
This choice will give us two estimates statistically independent of each other,
and, at the same time, by interleaving redshift bins, minimise evolutionary effects. 
This approach will also allow us to estimate random errors in the resulting
cosmological parameters by computing the standard error in the mean between the
two estimates. The resulting standard error can also be used as a fair estimate of
random errors in the two redshift bin case.

After computing the functions $\xi(r)$, we show in Fig.~\ref{sn3} the significance maps for 
the redshift bins 1-3 (upper panels) and 2-4 (lower panels). 
The percentage maps (left column) are shown for significant features averaged over two redshift bins.
When combining redshift bins 1-3 (upper panel) we observe that 18\%-26\% is the range of scales for 
the $\xi(r)$'s that are useful for our study while 19\%-31\% is the range of scales obtained for 
redshift bins 2-4 (lower panel). 
On the other hand, when analysing the averaged $S/N$ maps (right column) for significant features only,
the redshift bins 1-3 (upper panel) show a variation from 1.25 to 1.59 while the average over
redshifts 2-4 (lower panel) lies in the range 1.46 to 1.56.
It is clear that all cosmological pairs have some portion of the $\xi(r)$ useful for our study.

\begin{figure}
\centering
\includegraphics[width=9cm]{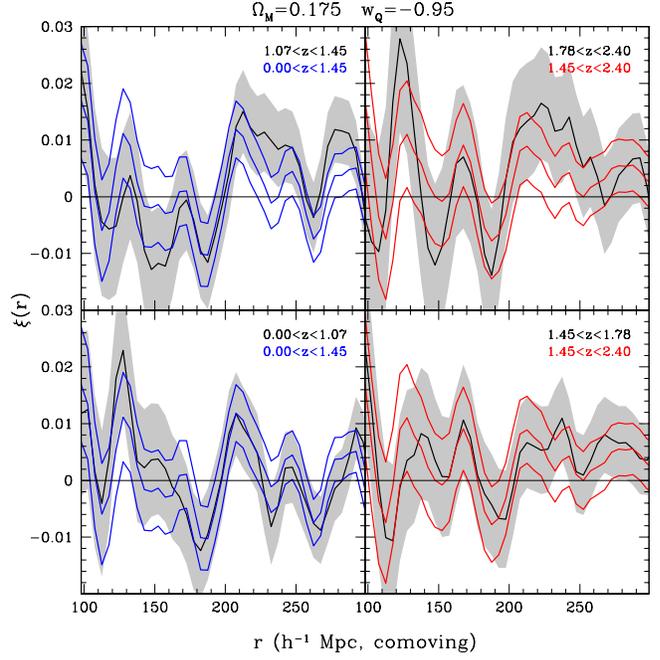}
\caption{
Comparison of the $\xi(r)$s between the two (blue and red curves) and four (grey areas) redshift bin cases 
for the preferred model obtained in the two redshift bin case. 
Error bars were computed using the bootstrap re-sampling technique.
Each panel quotes the corresponding redshift ranges.
}
\label{erosion}
\end{figure}

These $S/N$ maps span similar value ranges to those corresponding to the
two redshift bin case (see Fig.~\ref{sn}). Even when computing the $\xi(r)$s within narrower 
redshift bins produces larger error bars, on average, the $S/N$ show no degradation compared to
the previous case since some peaks and valleys in $\xi(r)$ are larger now. This could be a consequence
of less evolution within a single redshift bin. 
Under this assumption, some features in the $\xi(r)$ in the two redshift
bin case could have been eroded away by evolution. In Fig.~\ref{erosion} we observe a comparison
between the $\xi(r)$s obtained in the two (blue and red curves) and four (grey areas) redshift bins cases 
for the preferred cosmological model found in the previous subsection. Examples of our statement
are the regions with $200 \ \mpc \le r \le 250 \ \mpc$ in the two upper panels, and the
peak at $\sim 130 \ \mpc$ in the lower left panel of this Fig.

\begin{figure*}
\centering
{\includegraphics[width=8cm]{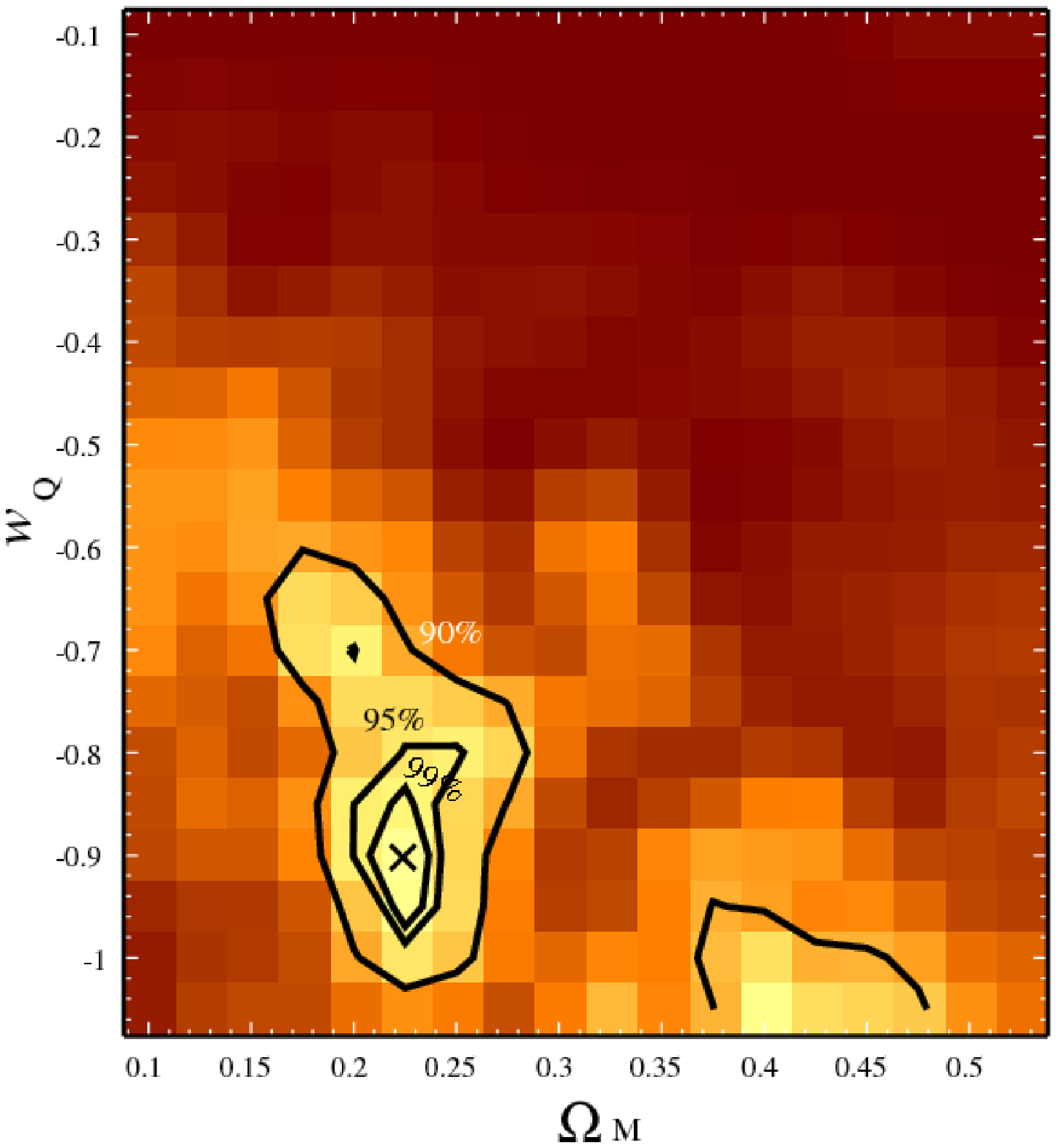}\includegraphics[width=8cm]{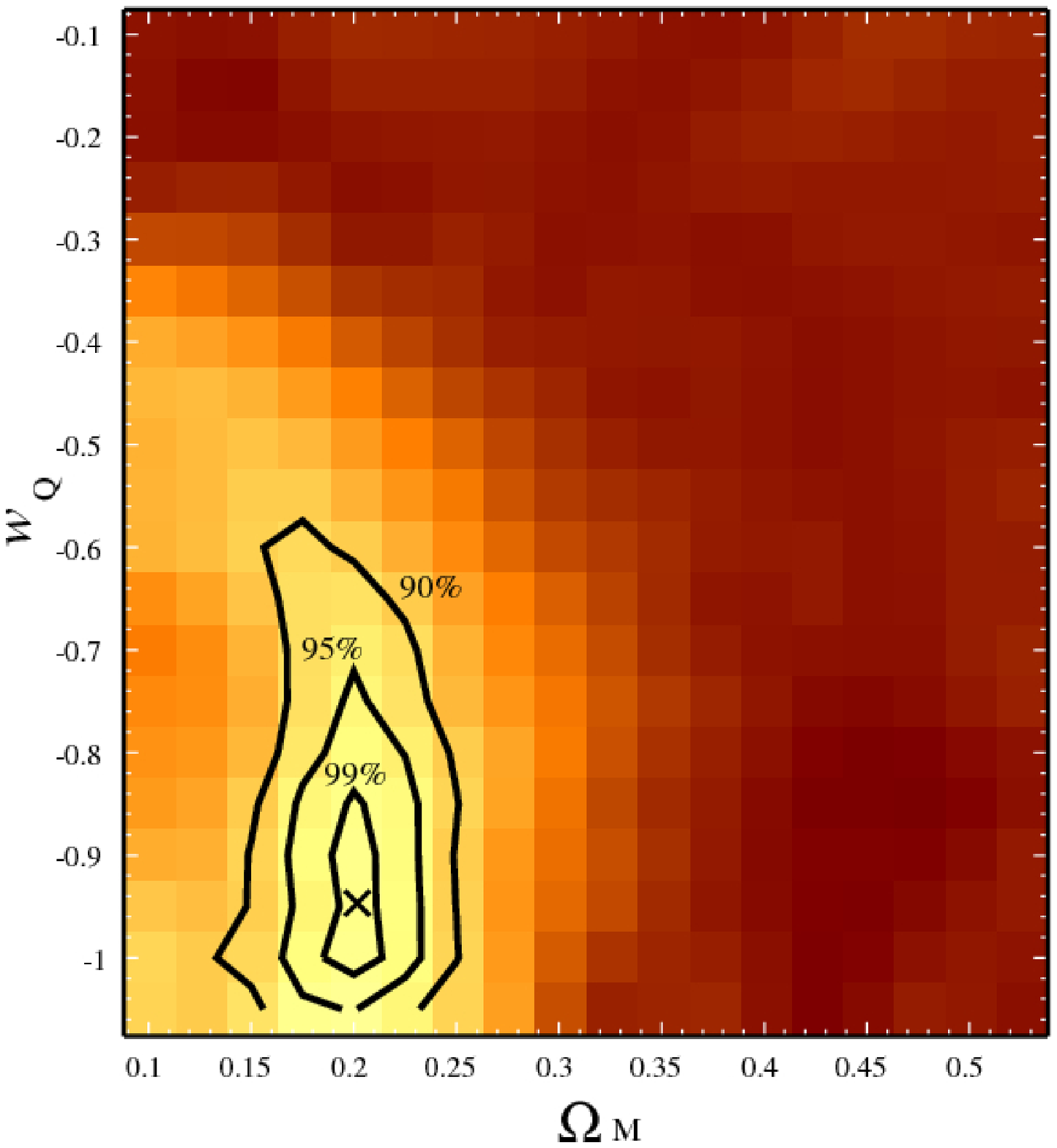}}
\caption{($\Omega_{\rm M}-w_{\rm Q}$) maps for the $CC^{(4z)}$ of the $\xi(r)$ for four redshift bins.
The left panel show the $CC^{(4z)}_{[1,3]}$, while the right panels show  $CC^{(4z)}_{[2,4]}$.
The minimum comoving distance for each map is $r_{min}=$110 $\mpc$. Each model is computed 
only for $S/N\ge $ 1.  
Higher values of the cross-correlation correspond to lighter colours. Labels in contour level show 
the percentage of the map outside the contours.
}
\label{cxi3}
\end{figure*}

Figure \ref{cxi3} shows the cross correlation maps corresponding to 
the calculation of equation \ref{cc} for redshift bins 1-3 ($CC_{[1,3]}^{(4z)}$, left panel)
and 2-4 ($CC_{[2,4]}^{(4z)}$, right panel). Each map is computed using $r_{min}=$ 110 $\mpc$.
The preferred model in the $CC_{[1,3]}^{(4z)}$ map is $\Omega_{\rm M}=0.225$ and $w_{\rm Q}=-0.90$.
On the other hand, the $CC_{[2,4]}^{(4z)}$ map better fits with $\Omega_{\rm M}=0.200$ and 
$w_{\rm Q}=-0.95$. We have checked that these results do not change when
varying $r_{min}$ from 120 to 190 $\mpc$.
From these results the mean cosmological values and their corresponding standard
errors are $\Omega_{\rm M}=0.21\pm0.02$ and $w_{\rm Q}=-0.93\pm0.04$.
We observe that this result is consistent within a $1 \ \sigma$ interval with the one obtained for the two 
redshift bin case.

In the light of our results, we consider as the most likely cosmological model to be the one obtained for 
the four redshift bin case since it is less affected by evolution effects while having similar
$S/N$ values to those obtained when using larger redshift bins.

%%%%%%%%%%%%%%%%%%%%%%%%%%%%%%%%%%%%%%%%%%%%%%%%%%%%%%%%%%%%%%%%%%%%%%%%%%%%%%%%%%%%%%%%
%%%%%%%%%%%%%%%%%%%%%%%%%%%%%%%%%%%%%%%%%%%%%%%%%%%%%%%%%%%%%%%%%%%%%%%%%%%%%%%%%%%%%%%%

%%%%%%%%%%%%%%%%%%%%%%%%%%%%%%%%%%%%%%%%%%%%%%%%%%%%%%%%%%%%%%%%%%%%%%%%%%%
%%%%%%%%%%%%%%%%%%%%%%%%%%%%%%%%%%%%%%%%%%%%%%%%%%%%%%%%%%%%%%%%%%%%%%%%%%%

\section{Summary and discussion}
\label{discussion}
In this work we have studied the evolution of the quasar two-point correlation function, $\xi(r)$, 
to constrain cosmological parameters for flat quintessential universes, namely the
matter adimensional density parameter $\Omega_{\rm M}$ and the quintessence equation
of state, $w_{\rm Q}$. 
For this purpose, we have drawn a large homogeneous quasar sample from the SDSS DR6, the
largest quasar catalogue available at present. The size and homogeneity of the
sample allow for highly reliable statistical studies of quasars.
As in previous studies of this aspect (e.g. \citealt{valli07,arielsan08,crocce08}), 
we have focused on analysing the position of
features in $\xi(r)$, at large comoving scales ($r>100\mpc$) well into the linear regime of structure 
formation. The key point of the analysis lies in the assumption that, for the proper cosmological
model, the comoving length scale of such features in $\xi(r)$ should remain fixed at different 
stages in the evolution of the universe. 
In this work, we do not assume any particular scale at which the features (peaks and valleys) 
should be found, but we only require consistency among different redshifts ranges, imposing
purely geometric constraints on the cosmological parameters.

The study is focused on two different complementary approaches: splitting the quasar redshift distribution
into two and four redshifts bins. The first analysis is performed in order to have reliable estimations of 
$\xi(r)$ at large scales (where the signal is expected to be low) while the second one is performed aiming 
to rule out a possible bias in the previous results due to an evolution of the quasar population within 
a particular redshift bin. 
The comparison among $\xi(r)$s of different redshift bins is carried out by means of the cross-correlation 
function ($CC$). This particular statistic has the advantage of quantifying the amount of overlap 
between two functions. In our case, the $CC$s are computed for modified $\xi(r)$s, i.e,  we have imposed
a significance criterion which keeps the $\xi(r)$s values when they satisfy the constraint 
$|\xi(r)/\sigma(r)|\ \ge \ 1$, otherwise are set to zero. 

With regard to the two redshift bin case, the 
most likely model is $\Omega_{\rm M}=0.18\pm0.02$ and $w_{\rm Q}=-0.95\pm0.04$. 
The coincidence among the results for the different reference scales ($r_{min}$) analysed is remarkable, 
implying that $\xi(r)$s for both redshift bins agree over a wide range of comoving scales, 
from $\sim 100$ to $\sim 300\mpc$.
From the four redshift bin case we obtain that the preferred cosmological model is
$\Omega_{\rm M}=0.21\pm0.02$ and $w_{\rm Q}=-0.93\pm0.04$.
When analysing the four redshift bin case, it becomes clear that the two redshift bin analysis is affected by
evolution. The main observable effect of evolution is that, on average, the amplitude of some features in 
the $\xi(r)$ has been diminished, and as a consequence, the best model has been slightly shifted towards
lower values of $\Omega_{\rm M}$.

The closest reference in the literature of an analysis similar to that performed in this work is
the study of $R02$. Some of the similarities with this work are: we both only use
a sample of quasars to constrain cosmological parameters, and neither of us search for a 
particular comoving scale.
Beyond quasar sample sizes, the methodology differences can be enumerated as follows:
\begin{itemize}
\item We extract information from the position of peaks and valleys in the $\xi(r)$s while $R02$ used 
peak information only.
\item We restrict the $\xi(r)$s with a significance criterion in order to avoid spurious detections while $R02$ 
used all the information without any restriction on $S/N$ ratio. 
\item The statistical tool proposed in this work intends to perform a full scale comparison 
among $\xi(r)$s, i.e,
to take into account more than one feature agreement among different redshift bins. 
On the other hand, $R02$ developed a method based on searching for the largest probability that the 
first local maximum in $\xi(r)$ above a given reference scale
occurs at the same comoving distance for consecutive redshift bins.
\item By construction, our method gives larger weights to stronger features while the 
probability method of $R02$ 
only considers the comoving length scale at which a peak is present.
\item We performed an analysis involving two and four redshift bins and found that evolution cannot be 
neglected. $R02$ performed their analysis using three redshift bins, extracting information from consecutive
redshift bins. Their approach can still be affected by evolution effects.
\end{itemize}
In their work, $R02$ found a matter density parameter in good agreement with our findings 
(see Table \ref{tabla}). Nevertheless, they cannot impose strong constraints on the quintessence parameter, 
finding only an upper limit ($w_{\rm Q}\le -0.50$).
Therefore, our results show that using the full large scale shape of
the $\xi(r)$s provides tighter constraints on cosmological parameters rather than 
basing the analysis on the coincidence of a single particular scale only.

%%%%%%%%%%%%%%%%%%%%%%%%%%%%%%%%%%%%%%%%%%%%%%%%%%%%%%%%%%%%%%%%%%%%%%%%%%%%%%%%%%%%%%%%
\begin{table*}
% \tabletypesize{\scriptsize}
\caption{Constraints on cosmological parameters $\Omega_{\rm M}$ and $w_{\rm Q}$ found in the
literature under the assumption of perfectly flat universes. 
The last column quotes the data sets used in each work.}
\label{tabla}
\begin{center}
\begin{tabular}{llll}
\hline\hline
Author & $\Omega_{\rm M}$ & $w_{\rm Q}$ & Data\\
\hline
\citet{roukema02}    & $0.25 \pm 0.10$   & $[-1.0,-0.5]$ & 10K 2dFQSO$^1$ \\
\citet{tegmark04}    & $0.329 \pm 0.074$ & $-0.92\pm0.30$ & WMAP$^2$ + SDSS MGS$^3$\\
\citet{eisenstein05} & $0.326 \pm 0.037$ & $-0.80\pm0.18$ & WMAP + SDSS MGS + LRG$^4$ \\
\citet{wang06}       & ...             & $-0.885^{+0.109~+0.206}_{-0.111~-0.227}$ & 
R04$^5$ WMAP3$^6$ + SNIa + LRG \\
\citet{wang06}       & ...             & $-0.999^{+0.082~+0.159}_{-0.083~-0.168}$ &  A06$^7$ WMAP3 + SNIa + LRG \\
\citet{arielsan06}   & $0.241\pm0.024$ & $-0.85^{+0.18}_{-0.17}$ & WMAP + 2dFGRS$^8$ \\
\citet{spergel07}    & $0.238\pm0.015$ & $-0.967\pm0.073$ & WMAP3 + SNLS$^9$ \\
This work            & $0.21\pm0.02$ & $-0.93\pm0.04$ & DR6 SDSS QSO \\
\hline
\hline
\end{tabular}
\parbox{14cm}{{\scriptsize
1:~Quasars from the 10K release of the 2 degree Field Quasar Redshift 
Survey, \citet{croom01}.\\
2:~First year Wilkinson Microwave Anisotropy Probe, \citet{spergel03}.\\
3:~Main Galaxy Sample, \citet{mgs}.\\
4:~Luminous Red Galaxies, \citet{eisenstein01}.\\
5:~\citet{riess04}.\\
6:~Third year Wilkinson Microwave Anisotropy Probe, \citet{spergel07}.\\
7:~\citet{astier06}.\\
8:~2 degree Field Galaxy Redshift Survey, \citet{2df}.\\
9:~Supernovae Legacy Survey, \citet{astier06}.}
}
\end{center}
\end{table*}
%%%%%%%%%%%%%%%%%%%%%%%%%%%%%%%%%%%%%%%%%%%%%%%%%%%%%%%%%%%%%%%%%%%%%%%%%%%%%%%%%%%%%%%%

From the comparison with results from a variety of data and techniques (see Table \ref{tabla}), 
we observe that our estimation of $\Omega_{\rm M}$ is in good agreement 
with $R02$, \citet{arielsan06} and \citet{spergel07}.
In addition, our result for $w_{\rm Q}$ agrees with all sources quoted in
Table \ref{tabla}, and has one of the smallest uncertainty intervals.

What are the possible implications of our
results in terms of structure formation compared to the initial
conditions established by the $\Lambda CDM$ concordance model? 
We observe a small discrepancy in the 
mean values obtained for the $\Omega_{\rm M}$ and $w_{\rm Q}$ 
compared with a standard universe with cosmological constant
($\Omega_{\rm M}=0.25$ and $w_{\rm Q}=-1.0$). Our results are
prone to prefer cosmologies with slightly lower values of $\Omega_{\rm M}$ 
and higher values of $w_{\rm Q}$ than those of the 
concordance model. Both tendencies observed in the parameters 
have similar consequences for structure formation. If all cosmological models were 
normalised to predict the observed amplitude of density fluctuations
at present, lower values of density would predict higher levels
of density fluctuations in the past (e.g. \citealt{spergel98}).
Under the same assumptions of density perturbation normalisation,
earlier cloud formation and higher core densities are observed if the dynamics of the dark energy
is enhanced, i.e. for models represented by an equation of state
parameter $w_{\rm Q}>-1$ \citep{bartelmann02,dolag04,maio06}. Hence, at a given 
mass and redshift our results would predict higher abundance
of more concentrated halos than in the model with the cosmological 
constant. 

Our results rely on the analysis of quasar clustering alone, 
demonstrating that it is plausible to put strong constraints on cosmological parameters
with only one suitable data set and a proper statistical treatment.
Since our results are independent of the cosmic microwave background, type Ia supernovae and galaxy data,
then large quasar samples in combination with all these data sets should imply a better 
determination of $\Omega_{\rm M}$ and $w_{\rm Q}$.

%%%%%%%%%%%%%%%%%%%%%%%%%%%%%%%%%%%%%%%%%%%%%%%%%%%%%%%%%%%%%%%%%%%%%%%%%%%%%%%%%%%%%%%%
%%%%%%%%%%%%%%%%%%%%%%%%%%%%%%%%%%%%%%%%%%%%%%%%%%%%%%%%%%%%%%%%%%%%%%%%%%%%%%%%%%%%%%%%
\section*{Acknowledgements}
We thank to the anonymous referee for important suggestions that have greatly improved
the original manuscript.
We also thank E.D. and Diego G. Lambas for carefully reading the manuscript and suggestions.
HJM acknowledges the support of a Young Researchers' grant from Agencia Nacional de 
Promoci\'on Cient\'\i fica y Tecnol\'ogica Argentina, PICT 2005/38087.
This work has been partially supported by grants from 
Consejo de Investigaciones Cient\'{\i}ficas y T\'ecnicas de la Rep\'ublica 
Argentina (CONICET).
Funding for the Sloan Digital Sky Survey (SDSS) has been provided by the 
Alfred P. Sloan 
Foundation, the Participating Institutions, the National Aeronautics and Space 
Administration, the National Science Foundation, the U.S. Department of Energy, 
the Japanese Monbukagakusho, and the Max Planck Society. The SDSS Web site is 
http://www.sdss.org/.
The SDSS is managed by the Astrophysical Research Consortium (ARC) for the 
Participating Institutions. The Participating Institutions are The University 
of Chicago, Fermilab, the Institute for Advanced Study, the Japan Participation 
Group, The Johns Hopkins University, the Korean Scientist Group, Los Alamos 
National Laboratory, the Max Planck Institut f\"ur Astronomie (MPIA), the 
Max Planck Institut f\"ur Astrophysik (MPA), New Mexico State University, 
University of Pittsburgh, University of Portsmouth, Princeton University, 
the United States Naval Observatory, and the University of Washington.
%%%%%%%%%%%%%%%%%%%%%%%%%%%%%%%%%%%%%%%%%%%%%%%%%%%%%%%%%%%%%%%%%%%%%%%%%%%%%%%%%%%%%%%%
%%%%%%%%%%%%%%%%%%%%%%%%%%%%%%%%%%%%%%%%%%%%%%%%%%%%%%%%%%%%%%%%%%%%%%%%%%%%%%%%%%%%%%%%

%%%%%%%%%%%%%%%%%%%%%%%%%%%%%%%%%%%%%%%%%%%%%%%%%%%%%%%%%%%%%%%%%%%%%%%%%%%%%%%%%%%%%%%%
%%%%%%%%%%%%%%%%%%%%%%%%%%%%%%%%%%%%%%%%%%%%%%%%%%%%%%%%%%%%%%%%%%%%%%%%%%%%%%%%%%%%%%%%
\end{document}